\documentclass[epj]{svjour}
\usepackage{amsmath,amssymb,epsfig,wrapfig} 

\usepackage[english]{babel}
\newcommand{\eqeqref}[1]{Eq.~\eqref{#1}}

\newcommand{\refref}[1]{Ref.~\cite{#1}}

\newcommand{\figref}[1]{Fig.~\ref{#1}}

\newcommand{\beq}{\begin{equation}}
\newcommand{\eeq}{\end{equation}}
\newcommand{\bea}{\begin{eqnarray}}
\newcommand{\beas}{\begin{eqnarray*}}
\newcommand{\beau}[1]{\begin{equation} \label{#1} \begin{array}{rcl}}
\newcommand{\eea}{\end{eqnarray}}
\newcommand{\eeas}{\end{eqnarray*}}
\newcommand{\eeau}{\end{array} \end{equation}}
\newcommand{\bay}{\begin{array}}
\newcommand{\eay}{\end{array}}
\newcommand{\bals}{\begin{align*}}
\newcommand{\eals}{\end{align*}}

\newcommand{\ds}{\displaystyle}

\newcommand{\lora}{{\longrightarrow}}

\newcommand{\vev}[1]{\langle #1 \rangle}

\newcommand{\PP}{{\mathcal P}}

\begin{document}
\title{ Space-time evolution of hadronization}
\author{
  Alberto Accardi
  \inst{1}
  \thanks{Based on a talk given at ``Hot Quarks 2006'', Villa Simius, Sardinia, Italy, May 15-20, 2006}
} 
\institute{Dept. of Physics and Astronomy, Iowa State U., Ames, IA
  50011, USA}
\date{}
\abstract{
Beside its intrinsic interest for the insights it can give into color
confinement, knowledge of the space-time evolution of hadronization is
very important for correctly interpreting jet-quenching data in heavy
ion collisions and extracting the properties of the produced
medium. On the experimental side, the cleanest environment to study the
space-time evolution of hadronization is semi-inclusive Deeply
Inelastic Scattering on nuclear
targets. On the theoretical side, 2 frameworks are presently competing
to explain the observed attenuation of hadron production: quark energy
loss (with hadron formation outside the nucleus) and nuclear
absorption (with hadronization starting inside the nucleus). 
I discuss recent observables and ideas which will help
to distinguish these 2 mechanisms and to measure the time scales of the
hadronization process.
 % end of PACS codes
} %end of abstract
\maketitle

\section{Introduction}
\label{sec:intro}

\begin{figure}[tb]
 \centerline{
  \includegraphics[height=4.3cm]{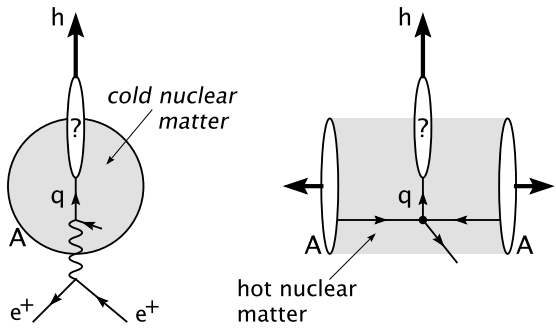}
 }
 \caption[]{Left: Lepton-nucleus scattering. The hadronizing quark travels
   through the target nucleus.
   Right: Nucleus-nucleus scattering. The hadronizing parton
   travels through the medium produced in the collisions.
 \label{fig:nuke}
 }
\end{figure}

One of the most striking experimental discoveries in the heavy-ion
program at the Relativistic Heavy Ion Collider (RHIC) has been the
the suppression of large transverse momentum hadron production in
nucleus-nucleus (A+A) collisions compared to proton-proton (p+p) collisions
\cite{RHIC}.
The observable of interest is the ratio of the hadron transverse momentum
($p_T$) spectrum in A+B collision in a given
centrality class (c.cl.), normalized to binary scaled p+p collisions 
by the inverse thickness function $T_{AB}$ and finally divided 
by the p+p spectrum:
\begin{equation}
  R_{AB} =  
    {\frac{1}{T_{AB}(c.cl.)} \frac{dN}{dp_T^2 dy}
       ^{\hspace{-0.3cm}A+B\rightarrow h+X}
       \hspace{-1.3cm} (c.cl.)} \hspace*{0.2cm} \Bigg/
    {\frac{d\sigma}{dp_T^2 dy}
       ^{\hspace{-0.3cm}p+p\rightarrow h+X}
       \hspace{-0.8cm} } \ .
  \label{CroninRatio}
\end{equation}
In the absence of nuclear effects, one would expect $R_{AB}=1$.
By comparing the measured $R_{AuAu} \approx 0.2$ in Au+Au collisions
to the mild deformation of $p_T$ spectra observed in deuteron-gold (d+Au)
collisions, one concludes that the large suppression of $R_{AuAu}$
is due to the hot and dense medium produced in Au+Au 
collision, also called ``hot nuclear matter'', see
\figref{fig:nuke}. This measurement is one of the keys to the claimed
discovery of the Quark-Gluon Plasma (QGP) at RHIC \cite{QGPdiscovery}.

The suppression of hadron production in A+A collisions, 
has been successfully described in terms of parton energy loss due to
medium-induced gluon bremsstrahlung, allowing
so-called ``jet tomography'' studies of the QGP
\cite{Gyulassy:2003mc,Vitev:2004bh}. 
However, this success has been recently questioned.
Gluon radiation off heavy quarks is expected on theoretical
grounds to be suppressed at small angles compared to light quarks,
implying a smaller suppression for $D$ and $B$ mesons than for $\pi$
mesons \cite{deadcone}. However, the measured 
suppression of single non-photonic electrons at RHIC \cite{singlee}, 
which are the decay product of $B$ and $D$ mesons, is of similar
magnitude  for pions contrary to theoretical expectations
\cite{HeavyQuarks2}.
The common assumption of neglecting elastic parton energy loss in
$R_{AA}$ computations has been recently reexamined \cite{Wicks},
but the effect seems insufficient to explain the data, at least within 
conventional schemes for treating the running coupling constant
\cite{Peshier}.
As further assumption that needs to be tested and will be addressed in
this paper, is that the quark
which fragments into the observed hadron traverses the whole medium and
hadronizes well outside it. If untrue, in-medium interactions and
screening of the hadronizing system would need to be accounted for.

Hadron suppression has also been observed in fixed target Deeply 
Inelastic lepton-nucleus Scattering (nDIS). 
In this case, the medium which induces the attenuation is the target
nucleus itself, also called ``cold nuclear matter'', see
Fig~\ref{fig:nuke}.  Experimental data 
are usually presented in terms of the ``multiplicity ratio''
\cite{EMC,HERMES1,HERMES2,HERMES3,JLAB} 
\begin{align}
  R_M^h(z_h) = \frac{1}{N_A^{DIS}}\frac{dN_A^h(z_h)}{dz_h} \Bigg{/}
    \frac{1}{N_D^{DIS}}\frac{dN_D^h(z_h)}{dz_h} ,\
    \label{MultiplicityRatio}	   
\end{align}
i.e., the single hadron multiplicity on a target of mass number $A$ 
normalized to the multiplicity on a deuteron target as a function of
the hadron's fractional energy $z_h=E_h/\nu$, where $\nu$ is 
the virtual photon energy. 
The ratios in the numerator and denominator cancel to a large extent
initial state effects like the modifications of parton distribution
functions due to
shadowing and EMC effects, exposing the nuclear modifications of the
fragmentation process: if we 
assume factorization formulae to be valid, we have at leading order 
$R_M \approx D^A_h / D^D_h$, i.e., the ratio of the fragmentation
functions (FF) in the nucleus A and in deuteron. 
If no nuclear effects modify the fragmentation process we
would expect $R_M \approx 1$. 
In fact, what is experimentally observed is 
hadron suppression 
in the $z_h=0.2-1$ and $\nu=2-20$ GeV range at HERMES
\cite{HERMES1,HERMES2,HERMES3}, and in the
$\nu=20-200$ GeV at the EMC experiment \cite{EMC}.  
The flavor dependence of the multiplicity ratio has also been
measured \cite{HERMES2,HERMES3}, showing suppression
for pions, kaons and antiprotons. 
Protons are enhanced at $z_h\lesssim 0.4$ and suppressed above: this
is a ``proton anomaly'' analogous to the ``baryon anomaly'' observed 
in p+A and A+A collisions \cite{RHIC,Cronin}.
Both the quenching and the enhancement increase with $A$. 
Data binned in  $\nu$, and in the photon
virtuality $Q^2$ are also available from HERMES. 
Very high-statistics measurements will be
available in the near future from the CLAS experiment at Jefferson
Labs \cite{JLAB}, with some preliminary results already presented
\cite{Brooks-JLAB}. 

The cleanest environment to address nuclear modifications of hadron
production is nuclear DIS: it allows to experimentally control
many kinematic variables; the nuclear medium (i.e., the nucleus
itself) is well known; the multiplicity in the final state is low,
allowing for precise measurements. Moreover, the nucleons act as
femtometer-scale detectors of the hadronizing quark, allowing to
experimentally study its space-time evolution into the observed hadron. 
Hadron suppression at HERMES and CLAS is of direct relevance to RHIC
physics. In both cases the
hadronizing quark has to traverse a length of (hot or cold) nuclear
matter of the size of a nuclear radius, see Fig~\ref{fig:nuke}. 
Moreover, in the HERMES experiment 
$\nu\approx 2-25$ GeV and $z_h\approx 0.2-1$, so that hadrons are
observed over an energy range $E_h = z_h\nu = 2-20$ GeV. 
(measurements at CLAS with $E_{beam} = 5,11$ GeV
will likewise explore the $E_h = 2-10$ GeV range).
For midrapidity hadron production in A+A collisions at RHIC 
$E_h \approx p_T \approx 2-20$ GeV,
roughly equal to the hadron energy range at HERMES.

Information about parton propagation in cold nuclear matter is needed as
an input for the interpretation of data in A+A collisions. In this case
one wants to use hadron suppression as a tool to extract the
properties of the hot QGP created in the collision. To this purpose we need 
to develop well calibrated computational tools to
relate the magnitude of hadron suppression to properties of the QGP
like its density and temperature. Assuming long lived partons, 
hadron suppression at RHIC would be attributed to parton energy loss, leading
to a medium temperature of $T\approx 400$ MeV \cite{Vitev:2004bh}, in
excess of the critical temperature $T_c\approx 170$ MeV 
for the QGP phase transition. If, on the contrary, hadronization started
on the nuclear radius scale or before, in-medium interactions should
also be accounted for, leading to a different, presumably lower $T$.
A precise knowledge of parton propagation and hadronization
mechanisms obtained from nDIS data is essential for testing and
calibrating our theoretical tools, and to determine the properties of
the QGP produced at RHIC.

\section{Formation time estimates}
\label{sec:formationtime}

The key quantity we need to investigate is the hadronization time scale. 
Since hadronization is a non perturbative process, 
one has to resort to phenomenological models to describe it
\cite{Wang,Arleo,Bialasgyulassy,Accardi:2002tv,AGMP05,Kopeliovich,Falteretal04}.   
However, a few features are expected on general grounds.
Due to color confinement, the struck quark must neutralize its color
at some stage, say by picking up an antiquark from the
vacuum or the surrounding medium. I call this color neutral $q\bar q$ pair a
``prehadron'' $h_*$, and the time for its formation the ``prehadron
formation time''  $t_*$ (some authors prefer to call it the
  ``production'' time). This is a relevant time scale since
gluon bremsstrahlung off the struck quark stops after color
neutralization; moreover, the prehadron quickly develops a
cross section of the order of the hadronic one, leading to its nuclear
absorption. Subsequently, the prehadron wave function
collapses on the observed hadron $h$ wave function, and the
corresponding time is called ``hadron formation time'' $t_h$. 
A final caveat: it is difficult to
rigorously define the concept of formation time in field theory, so that
in the following discussions it is used as a working tool. 

\begin{figure*}[tb]
  \vspace*{-.2cm}
  \begin{center}
  \includegraphics[height=5.6cm,origin=c]{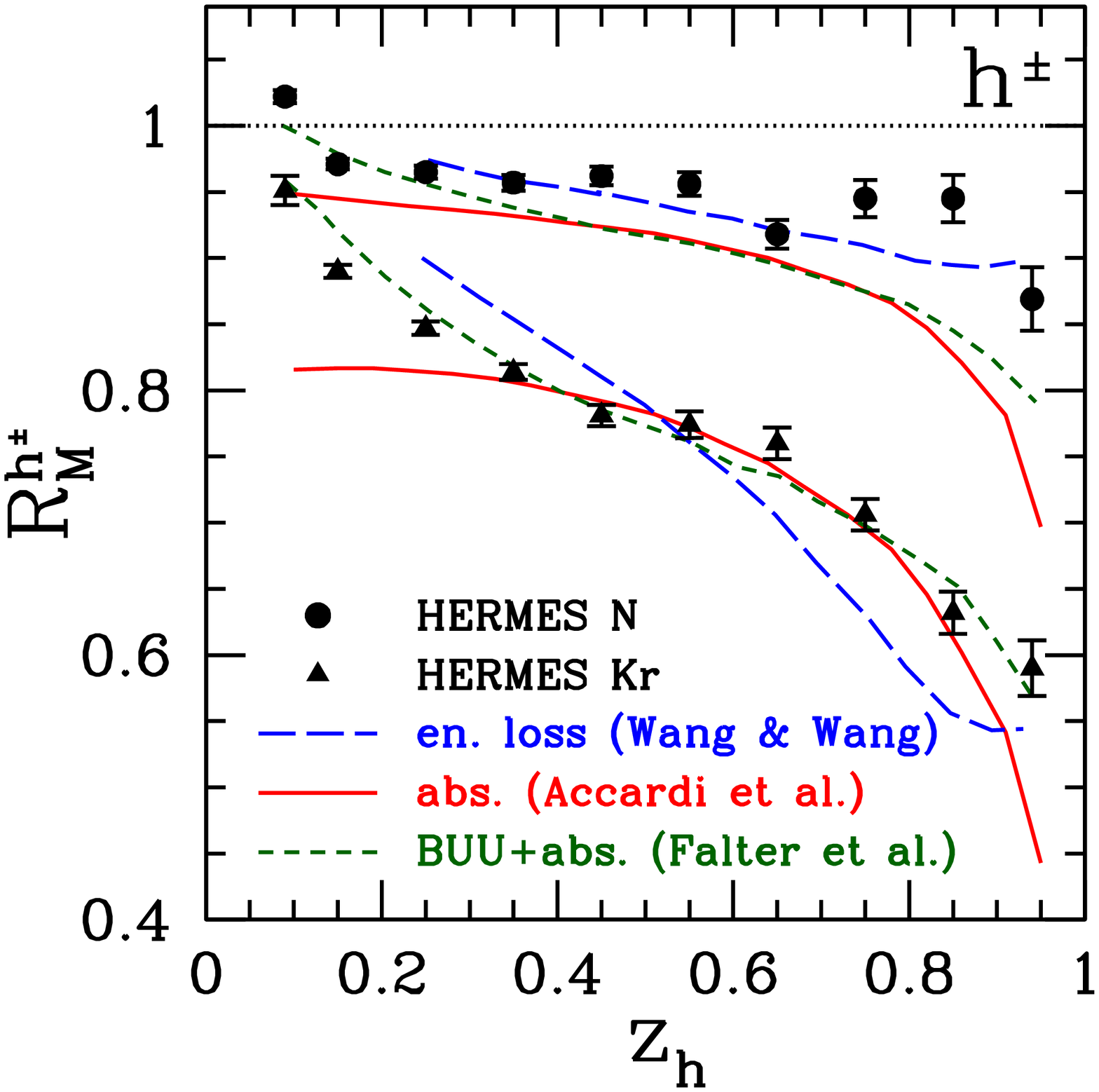}
  \includegraphics[height=5.6cm,origin=c]{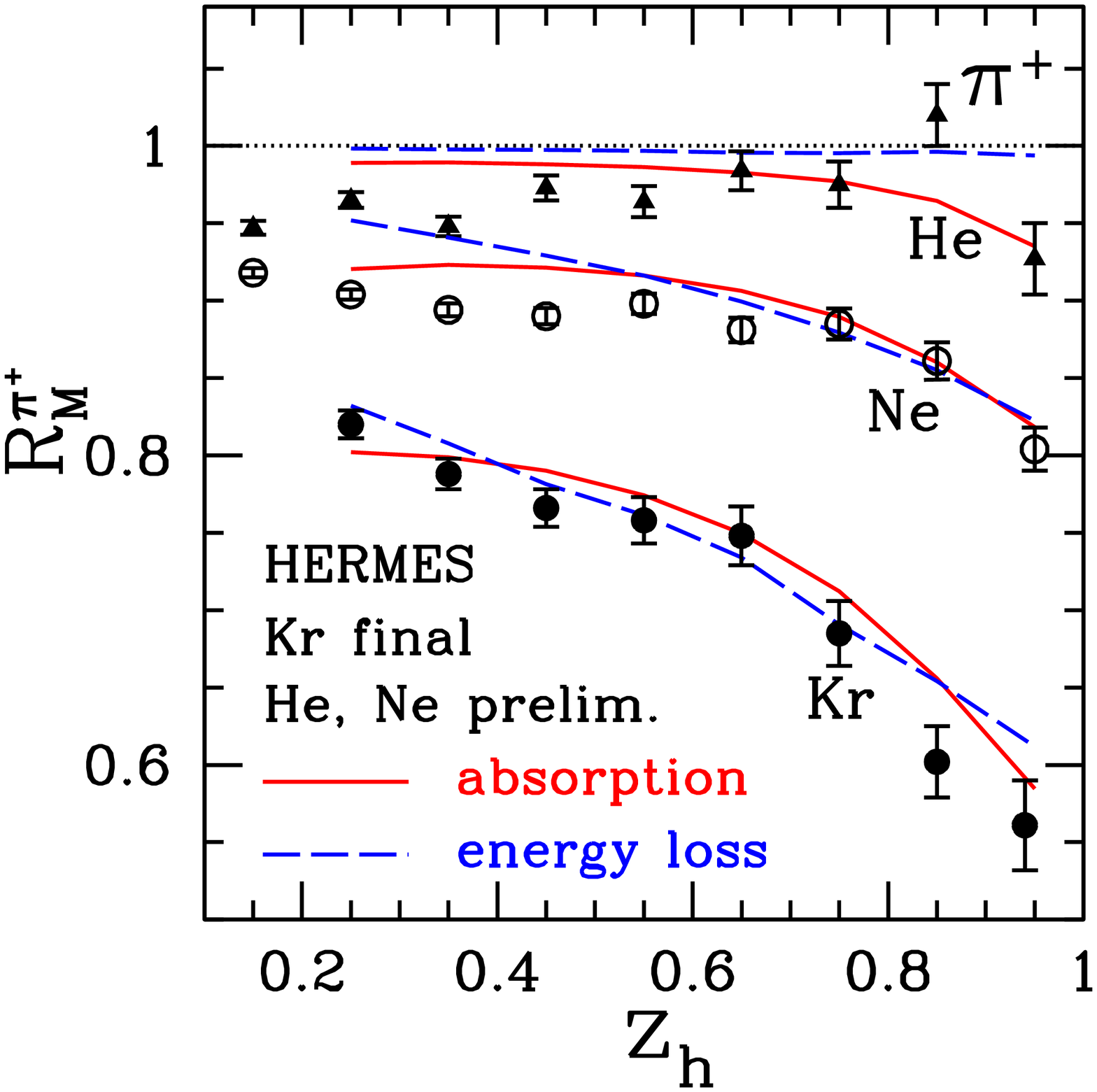}
  \includegraphics[height=5.55cm,width=5.9cm,origin=c]{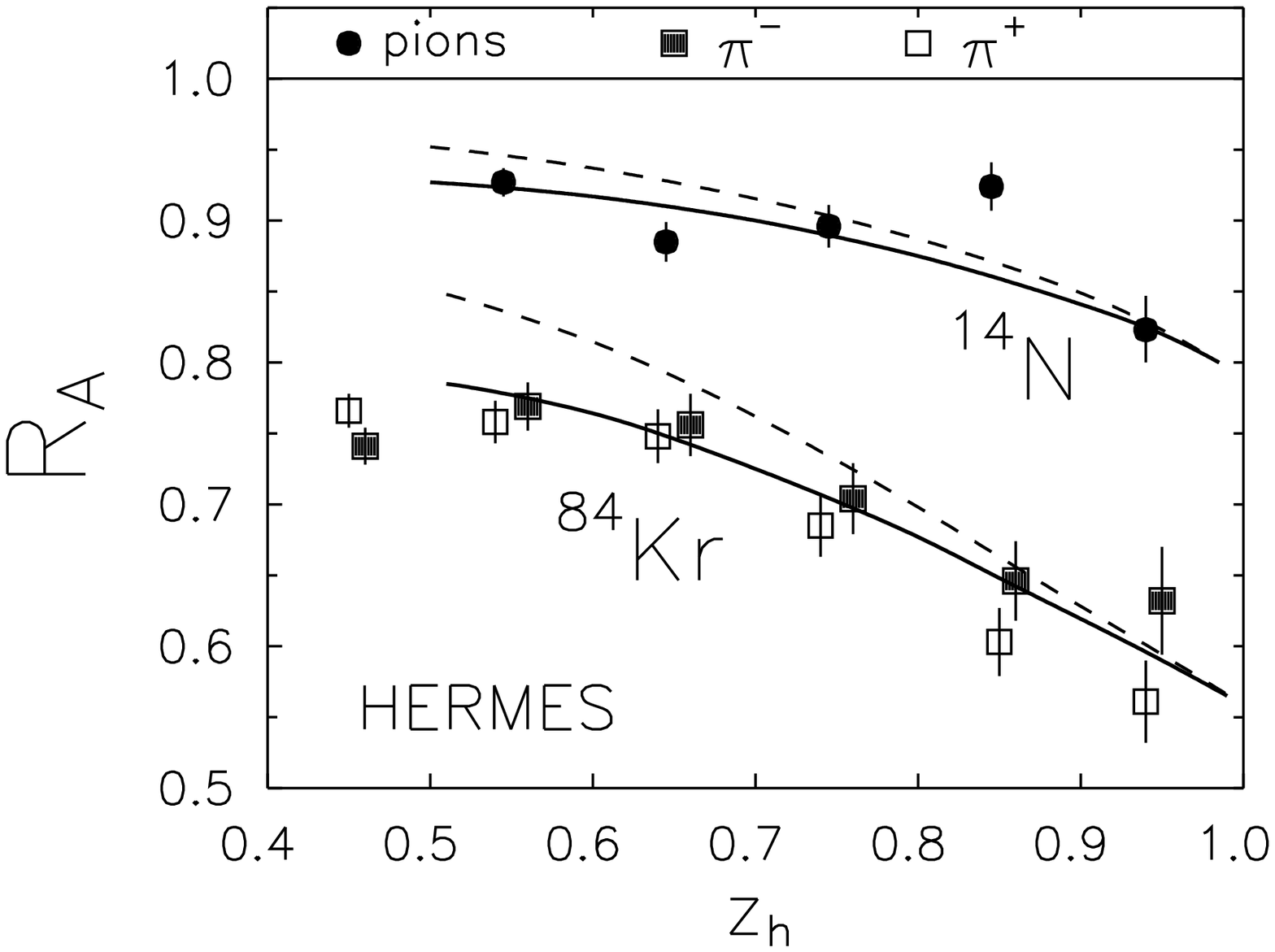}
  \end{center}
 \vspace*{-0.3cm}
 \caption[]{
   Left:Multiplicity ratio for $h^\pm$ at HERMES \cite{HERMES2} 
   compared to the energy loss model of \refref{Wang} and the 
   absorption models of \refref{AGMP05,Falteretal04}.
   Center: pure absorption model \cite{AGMP05,Accardi:2005mm} (solid)
   and energy loss model \cite{Accardi:2005mm,Arleo} (dashed). 
   Right: Color dipole model from Ref.~\cite{Kopeliovich} (dashed: absorption only,
   solid: absorption and induced energy loss).
   Data are from Ref.~\cite{HERMES1,HERMES2,HERMES3}.
 \label{fig:AAvsAGMP}
 \label{fig:AGMP_AA_Kop}
 }
\end{figure*}

\subsection{Long formation time: energy loss models}

The average hadron formation time $\vev{t_h}$ can be considered as the
time for the 
struck partons to build up its color field and to develop the
hadronic wave function \cite{Dokshitzer:1991wu}. In the 
hadron rest frame this time is related to the hadron radius, and in the
laboratory frame it is boosted to:
\begin{align} 
  \vev{t_h} \propto R_h \frac{E_h}{m_h} = R_h \frac{z \nu}{m_h}
 \label{eq:wangest}
\end{align}
For a 10 GeV pion at HERMES, we have $\vev{t_h} \approx 50$ fm $\gg R_A$.
Note also that the
scale for hadron formation is set by $\kappa_h = m_h/R_h \approx 0.2$
GeV/fm. This estimate is used in energy loss models \cite{Wang,Arleo}
to assume that hadronization starts outside the medium with a
decreased parton energy due to multiple parton
scatterings and induced gluon bremsstrahlung. These models are fairly
successful in describing $R_M$ at HERMES, see
Fig. \ref{fig:AGMP_AA_Kop}. 

More in detail, the Ref.~\cite{Wang} computes
parton rescatterings and gluon radiation
in pQCD including Feynman diagrams up to twist-4.
Fragmentation of both the struck
quark and the radiated gluon is included. The modification of the FF
depends on 1 parameter, the strength of parton-parton
correlations in the nucleus. The modified FF so obtained
can be modeled to a good accuracy by shifting $z_h$ in the
leading-twist fragmentation function:
\begin{align}
  \tilde D (z_h) \, \lora \, \frac{1}{1-\Delta z_h} 
    D\Big(\frac{z_h}{1-\Delta z_h} \Big)  
 \label{eq:zshift}
\end{align}
where $\Delta z_h = \Delta E_q / \nu$ is the quark's fractional energy
loss, and $\Delta E_q \approx 0.6 \vev{z_g}$ with $\vev{z_g}$ the
average fractional energy carried away by the radiated gluon.

In \refref{Arleo} the parton energy loss is treated in the BDMS
formalism \cite{Baier:2001yt}, which takes into account medium-induced 
multiple soft gluon emission and fluctuations in the energy loss.
Modified FF are computed as the average
of the $z_h$-shifted FF in \eqeqref{eq:zshift}:
\begin{align*} 
  \tilde D_f^h (z_h) = & \int\limits_0^{(1-z)}\!\!\!\! d\Delta z\; 
    \PP(\Delta z\,;\hat q,L_q) \frac{1}{1-\Delta z}
    D_f^h(\frac{z_h}{1-\Delta z}) \ .
\end{align*}
The ``quenching weight'' $\PP(\Delta z)$ 
is the probability distribution of a fractional energy loss $\Delta
z=\Delta E_q/\nu$ \cite{Baier:2001yt,SW03}, and $L_q$ the quark's
in-medium path length. The medium is characterized by the transport
coefficient $\hat q$, which measures the average momentum transfer
per unit path-length from the medium to the parton.
When also taking into account a realistic medium geometry and 
finite medium size corrections to the quenching weight, the model 
\cite{Accardi:2005mm}can well describe HERMES data, see
Fig.~\ref{fig:AAvsAGMP}. 

\subsection{Short formation time: nuclear absorption models}
\label{sec:short}

A successful non perturbative model of hadronization is the Lund string
model \cite{Andersson:1983ia}. The confined color field
stretching from the 
struck quark to the rest of the nucleus is modeled as a
string of tension $\kappa_{str} \approx 1$ GeV/fm.
Prehadron formation is identified with the $q\bar q$ pair production
which breaks the string in smaller pieces \cite{Bialasgyulassy}.
Hadrons are formed when a quark and an antiquark at the endpoint of a string
fragment meet. Average formation times can
be analytically computed \cite{Bialasgyulassy,Accardi:2002tv,AGMP05}:
\begin{align}
  \vev{t_*} & = f(z_h) (1-z_h) \frac{z_h \nu}{\kappa_{str}} \\
  \vev{t_h} & = \vev{t_*} + \frac{z_h \nu}{\kappa_{str}} \ .
 \label{eq:lundest}
\end{align}
The factor $z_h \nu$ can be understood as a Lorentz boost factor.
The $(1-z_h)$ factor is due to energy conservation: a high-$z_h$ hadron
carries away an energy $z_h\nu$; the string remainder has a small
energy $\epsilon=(1-z_h)\nu$ and cannot stretch farther than
$L=\epsilon/\kappa_{str}$. Thus the string breaking must occur on a
time scale proportional to $1-z_h$. The function $f(z_h)$ 
is only a small deformation of $\vev{t_*}$. 
At HERMES $\vev{t_*} \approx
4$ fm $< R_A$ and $\vev{t_h} \approx 6-10$ fm $\gtrsim R_A$. The hadron is
typically formed at the periphery or outside the nucleus so that its
interaction with the medium is negligible. 
However, the prehadron is formed well inside and can start interacting
with the nucleus.
The nuclear absorption model of \refref{AGMP05} can successfully 
explain $R_M$ measurements at HERMES in terms of 
prehadron-nucleon inelastic scatterings with the
above formation times estimate, see Fig.~\ref{fig:AAvsAGMP}.
The prehadron-nucleon inelastic cross-section is $\sigma_*(\nu)
= 0.35 \, \sigma_h(\nu)$ proportional to the experimental hadron-nucleon cross
section $\sigma_h$. The proportionality factor is fitted to 
$\pi^+$ production data on a Kr target \cite{HERMES2}.
The prehadron survival probability $S_*$ is computed in terms of transport
equations. Neglecting hadron absorption, 
\begin{align} 
  S_* = & \int d^2b\,dy\,\rho_A(b,y)  \nonumber  \times 
    \int\limits_{y}^{\infty}dx\, 
    \frac{e^{-\frac{x-y}{\left< t_* \right>}}}
    {\left<t_*\right>}e^{-\sigma_*\int\limits_x^{\infty}dsA\rho_A(b,s)} 
    \ ,
\end{align}
where $(b,y)$ is the $\gamma^*$-q interaction point, 
$\rho_A$ is the nuclear density, and 
one can recognize exponential probability
distributions for prehadron and hadron formation.

In Ref.~\cite{Kopeliovich} the formation of a leading hadron ($z_h
\gtrsim 0.5$) is described in a pQCD
inspired model. The struck quark radiates a gluon 
according to the Bethe-Heitler radiation spectrum. 
The gluon then splits into a $q\bar q$ pair, and the $\bar q$
recombines with the struck $q$ to form the leading
prehadron. Medium interaction
and evolution of the prehadron into the observed hadron is 
computed in a path-integral formalism for dipole propagation.
The prehadron formation time is identified with the time at which the
gluon becomes decoherent with the struck quark.
The probability distribution in the prehadron formation time can be
computed, and the average $\vev{t_*}$ is
\begin{align}
  \vev{t_*} \propto (1-z_h) \frac{z_h \nu}{Q^2} \ .
\end{align}
The scale is set by $\kappa_{dip} = Q^2 \approx 10$ GeV/fm at HERMES, 
and $\vev{t_*} \lesssim 5$ fm at $z_h>0.5$. This model can
successfully describe leading hadron suppression, see
Fig.~\ref{fig:AAvsAGMP} right. 

In \refref{Falteretal04} a different space-time picture of hadronization
is advocated. Prehadrons are formed at $t_*=0$, and
hadrons are formed at $t_h=(E_h/m_h) \tau_0$ with $\tau_0=0.5$ fm.
The leading prehadron interacts with the medium with a reduced
hadronic cross-section. Subleading prehadrons do not
interact with the medium until hadron formation.
This picture is then embedded in a Monte Carlo transport model.
A good description of HERMES data can
be achieved, see Fig.~\ref{fig:AAvsAGMP}. 

\section{Can we distinguish energy loss from hadron absorption?}
\label{sec:distinguish}

Most of the difference in the time
estimates quoted in the previous Section lies in the 
different scale $\kappa$ used. E.g., $\kappa_h \approx 0.2
\kappa_{str}$ leads to the rather large $\vev{t_h} \approx 50$ fm quoted in
the energy loss model estimate instead of $\vev{t_h} \approx 10$ fm 
quoted in the Lund
model estimate. In the second case there would be no justification for
neglecting the interactions of the forming hadron field with the
nucleus.
As the choice of the scale $\kappa$ is a debatable and model-dependent
matter \cite{Kopeliovich,Wang:2003aw},
it is very important to look for observables which are able to
distinguish energy loss models and absorption models, or to directly
detect in-medium hadronization effects. 

\begin{figure}[tb]
  \vspace*{0cm}
  \centerline{
  \includegraphics[height=0.47\linewidth]{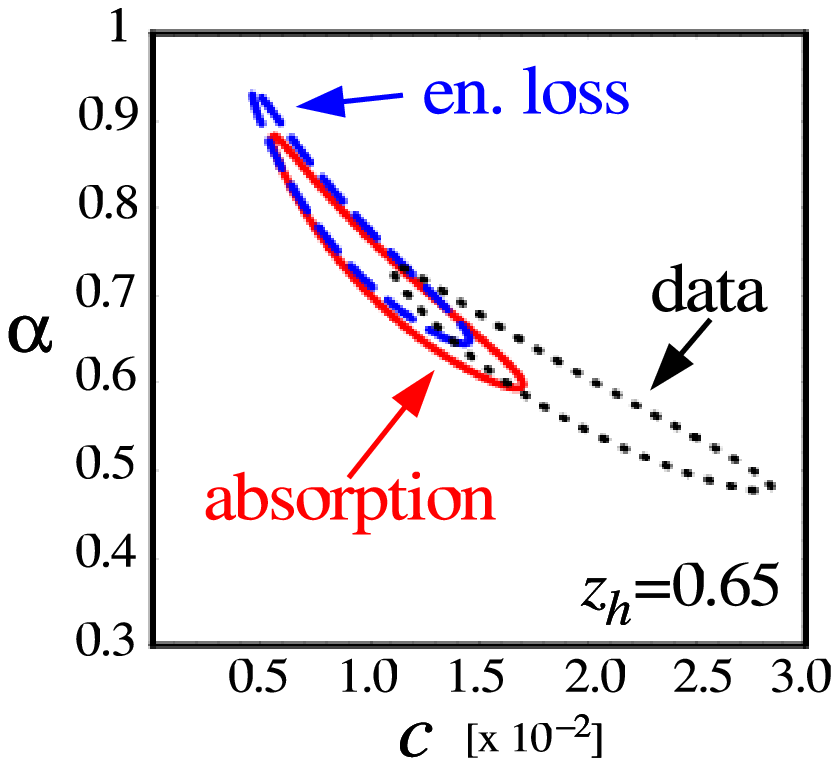}
  \includegraphics[height=0.47\linewidth]{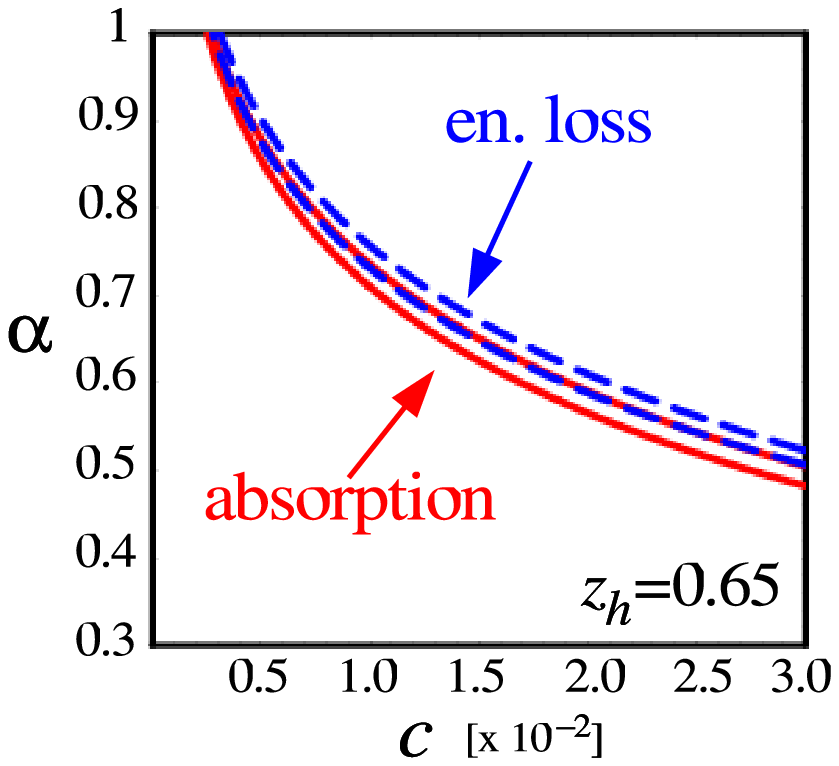} 
  }
  \vspace*{0cm}
 \caption[]{
  Left: results of the $R_M=cA^\alpha$ fit for \{He,N,Ne,Kr\} at
  z=0.65 (solid: absorption; dashed: energy loss; dotted: data
  \cite{HERMES2}). 
  Right: computations including only heavy nuclei \{Kr,Sn,Xe,W,Au,Pb\}.
 }
 \label{fig:RMplus}
\end{figure}

\subsection{Mass number dependence}
\label{sec:Adep}

%\noindent 
In first approximation, one expects $1-R^h_M \propto
A^{2/3}$ in energy loss models because the average energy loss
$\Delta E_q \propto \vev{L_q^2} \propto A^{2/3}$, due to the
Landau-Pomeranchuk-Migdal interference in QCD \cite{ptbroad}.
On the other hand, in absorption models the survival probability is
proportional to the amount of traversed matter, so that $1-R^h_M
\propto \vev{L_A} \propto A^{1/3}$. Therefore a
simple analysis of the $A$-dependence of $R_M^h$ should clearly signal
which model is correct. 

This argument fails for absorption models \cite{AGMP05,Blok:2005vi}. 
If the 
prehadron were produced always at the $\gamma^*$-quark interaction
point (i.e., $t_*=0$) then $R_M = c \,A^{1/3}$ at all
orders in $A^{1/3}$. However, if we allow for a nonzero $\vev{t_*}$, 
its dimension must be neutralized by the nuclear radius $R_A$,
introducing extra powers of $A^{1/3}$. Quite generally, if the
probability distribution for the prehadron formation 
length is finite at zero formation length, then $R_M^h \propto A^{2/3} +
O(A)$, the same power found in energy loss models. 

Then, we can study the {\em breaking} of the $A^{2/3}$ law. To this
purpose, it was proposed in \cite{AGMP05} to select a set of targets
$\{A_1,A_2,\ldots,A_n\}$, fix the $z$ bin, 
and perform a fit of the form $1-R^h_M(z) =
c(z) A^{\alpha(z)}$ with $c$ and $\alpha$ free
parameters. Results are presented in terms of $2\sigma$
confidence contours in the $(c,\alpha)$ plane.
As shown in \figref{fig:RMplus} left, energy loss
\cite{Accardi:2005mm} and absorption models
\cite{AGMP05,Accardi:2005mm} 
are indistinguishable. The same holds true
for all $z$ bins. Experimental data are described by an $A^\alpha$
power law with $\alpha=0.61 \pm 0.14$, compatible with $\alpha=2/3$
but excluding $\alpha=1/3$.
Increasing the number of targets and the span in
atomic number does not help in separating the 2 models, either, but clearly
shows a non negligible breaking of the $A^{2/3}$ law at $A \gtrsim
80$ \cite{AGMP05,Accardi:2005mm}, see \figref{fig:RMplus} right.

\begin{figure*}[tb]%\sidecaption
  \centerline{\parbox{14.3cm}{
    \vspace*{-0.1cm}
    \includegraphics[height=7cm,origin=c]{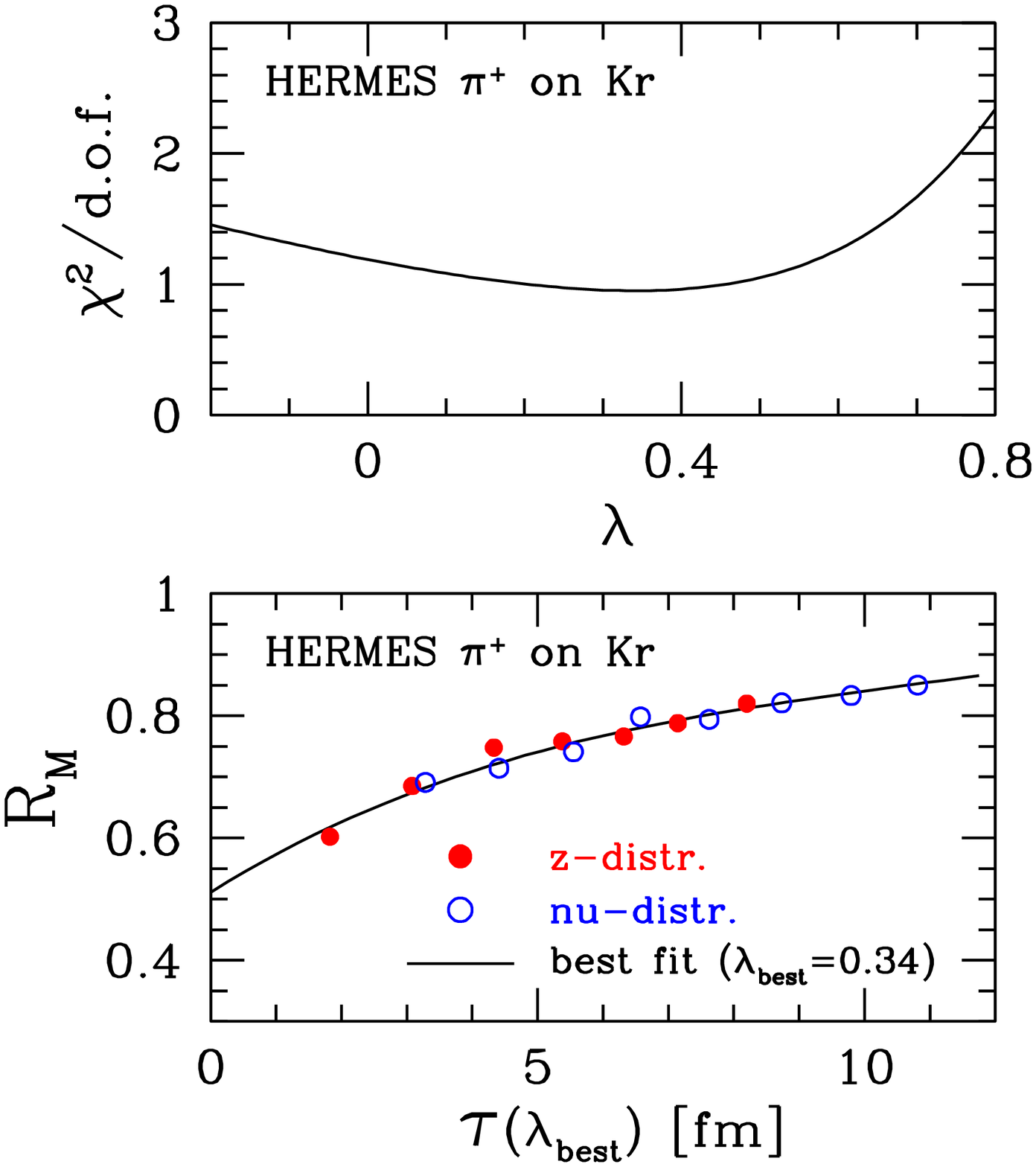}
    \hspace*{-.9cm}
    \includegraphics[height=7cm,origin=t]{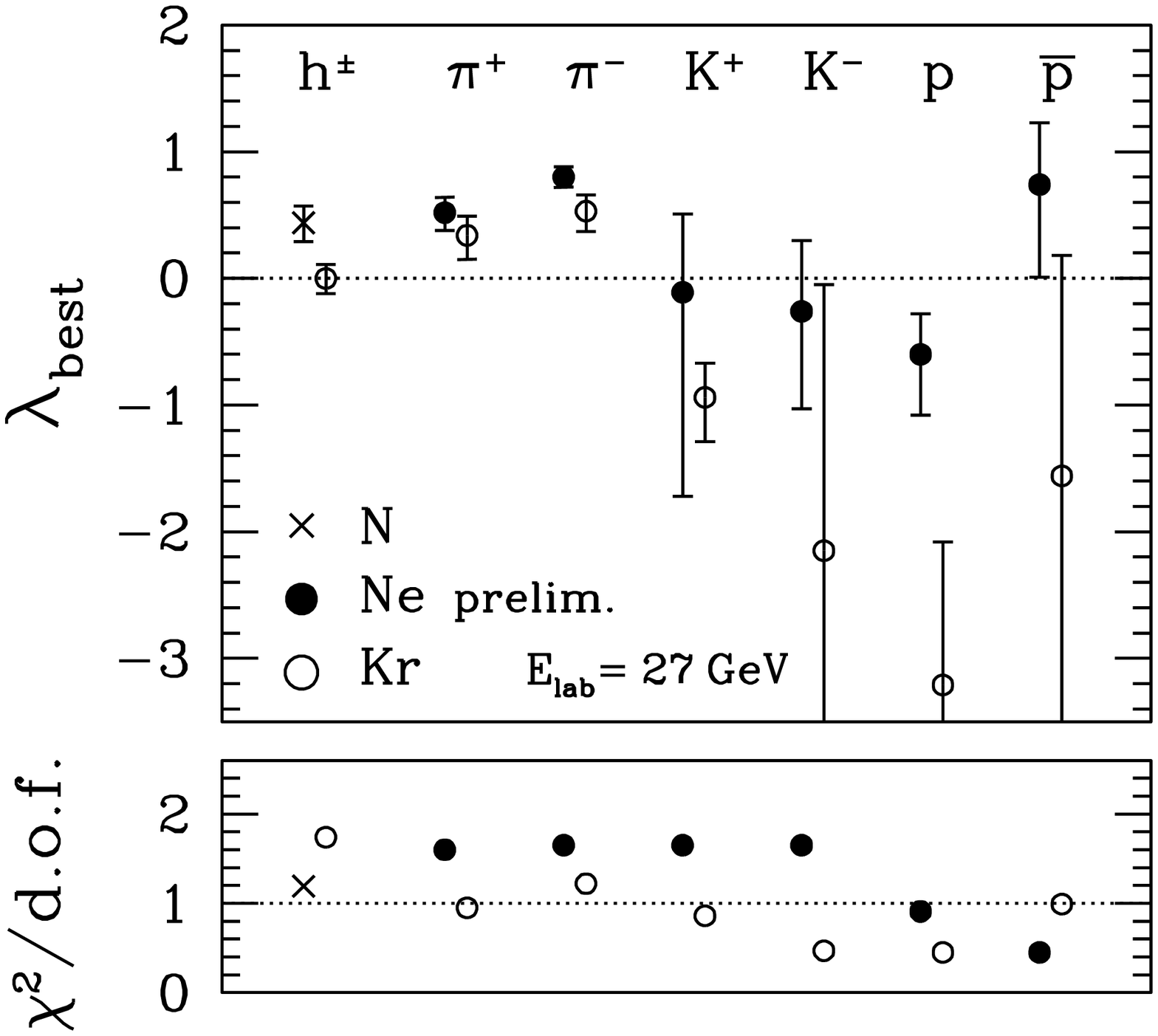} 
  }}
 \caption[]{Left: An example of the fit procedure described in
   Section~\ref{sec:scaling}  applied to HERMES data for $\pi^+$
   production on a Kr target \cite{HERMES2}. Upper panel: $\chi^2$ as
   a function of $\lambda$. Lower panel: $R_M(\tau)$ with $\tau$
   computed at $\lambda_{\rm best}=0.34$. Right: The scaling exponent
   $\lambda_{\rm best}$ extracted from HERMES data on charged and
   identified hadrons at $E_{lab}=27$ GeV
   \cite{HERMES1,HERMES2,HERMES3} (only statistical errors included in
   the fit). Error bars correspond to 1 standard deviation. The bottom
   panel shows the $\chi^2$ per degree of freedom. %\vskip-2.2cm 
   }
 \end{figure*}

\subsection{Formation time scaling}
\label{sec:scaling}

In Ref.~\cite{Accardi06} I conjecture that $R_M$ should not
depend on $z_h$ and $\nu$ separately but should
depend on a combination of them:
\begin{align}
  R_M = R_M\big[\tau(z_h,\nu)\big] \ ,
 \label{eq:RMscaling}
\end{align}
where the scaling variable $\tau$ is defined as
\begin{align}
  \tau & = C\, z_h^\lambda (1-z_h) \nu \ .
 \label{eq:scalingvar}
\end{align}
The scaling exponent $\lambda$ is introduced as a way of
approximating and summarizing the scaling behavior of experimental
data and theoretical models. It can be obtained by a best fit
analysis of data or theoretical computations.
The proportionality constant $C$ cannot
be determined by the fit. 
A possible scaling of $R_M$ with $Q^2$ is not considered in this
analysis because of its model dependence, and because of 
the mild dependence of HERMES data on $Q^2$.
As discussed below, the proposed functional
form of $\tau$, Eq.~\eqref{eq:scalingvar}, is flexible enough to
encompass both absorption models 
and energy loss models. The 2 classes of models are distinguished by 
the value of the scaling exponent: a positive $\lambda \gneqq 0$ is
characteristic of absorption models, while a negative $\lambda \lesssim
0$ is characteristic of energy loss models. Thus, the exponent
$\lambda$ extracted from experimental data can identify in the
experimental data the leading mechanism for hadron suppression in
nDIS.

The scaling of $R_M$ is quite natural in the context of hadron
absorption models
\cite{Bialasgyulassy,Accardi:2002tv,AGMP05,Kopeliovich,Falteretal04}. Indeed,
prehadron absorption depends on the
in-medium prehadron path length, which depends solely on
the prehadron formation time $\vev{t_*}$ as long as $\vev{t_*}
\lesssim R_A$. As argued in 
Section~\ref{sec:formationtime}, $\vev{t_*} \propto f(z_h) (1-z_h) z_h \nu$
which is well described by the proposed scaling variable $\tau$ with
$\lambda>0$. E.g., in the  Lund model $\lambda \approx 0.7$. 

In energy loss models \cite{Wang,Arleo,Accardi:2005mm} the scaling is
less obvious. For the purpose of discussing the 
scaling properties of $R_M$, we can consider the model of 
Ref.~\cite{Arleo,Accardi:2005mm} and neglect finite medium size
corrections and finite quark energy corrections.
If we further neglect energy loss fluctuations, we can 
approximate $R_M\approx \widetilde D_A(z_h)/D(z_h)$ and obtain
\begin{align*}
  R_M \approx  \frac{1}{1-\vev{\epsilon}/\nu} 
    D \Big( \frac{z_h}{1-\vev{\epsilon}/\nu} \Big) 
    \, \big[ D(z_h) \big]^{-1} \ ,
% \label{eq:approxRM}
\end{align*}
where the average energy loss $\vev{\epsilon} = \int_0^{(1-z_h)\nu}
d\epsilon\,\epsilon\,{\cal P}(\epsilon)$ \, 
$/ \int_0^{(1-z_h)\nu}
d\epsilon\,{\cal P}(\epsilon)= f[(1-z_h)\nu]$ is a
function of the energy $(1-z_h)\nu$ not carried away by the observed
hadron.
Next, we can approximate the FF using the parametrization of
\refref{KKP00} at $Q^2=2$ GeV$^2$ and obtain
\begin{align*}
  R_M \approx 
    \frac{1}{\Big( 1 - \ds\frac{1}{\nu} f[(1-z_h)\nu] \Big)^{\alpha+\beta+1}}  
    \left( 1 - \frac{f[(1-z_h)\nu]}{(1-z_h)\nu} \right)^\beta 
\end{align*}
This shows an approximate scaling with $(1-z_h)\nu$, which
implies scaling of $R_M$ with respect to $\tau$ with 
$\lambda \approx 0$. A similar argument holds for the model of
\cite{Wang}. When performing the scaling analysis of the full models
one finds in general $\lambda \lesssim 0$ \cite{Accardi06}.

The HERMES experiment measures $R_M$ binned in $z_h$ and integrated
over $\nu$ and $Q^2$ (``$z_h$ distributions'') or binned in $\nu$ and
integrated over $z_h$ and $Q^2$ (``$\nu$ distributions''). 
\eqeqref{eq:RMscaling} is fitted to the combined $z_h$-
and $\nu$-distributions, and the scaling exponent $\lambda$ is
determined by $\chi^2$ minimization. An example of this procedure is
illustrated in Fig.~\ref{fig:lambdafit}. Details can be found in
Ref.~\cite{Accardi06}. 

The scaling exponents $\lambda_{\rm best}$ extracted from HERMES data
at $E_{lab}=27$ GeV \cite{HERMES1,HERMES2,HERMES3} for different
hadron flavors produced on N, Ne and Kr targets  
are shown in Fig.~\ref{fig:lambdafit}. In all cases $\chi^2/{\rm
  d.o.f.} \lesssim 1.6$, which proves that $R_M$ scales with $\tau$.
The central result of this analysis is that pion data 
exhibit a clear $\lambda_{\rm best} \approx 0.4 \gneqq 0$. As
discussed, this shows the dominance of the prehadron absorption
mechanism as opposed to the energy loss mechanism, or in other words
is a signal of in-medium prehadron formation, with formation times
$\vev{t_*}\lesssim R_A$. 

\subsection{$p_T$ broadening}

The scaling analysis just described 
gives only indirect evidence for a short formation time,
and cannot measure its absolute scale. 
An observable which is more
directly related to the prehadron formation time is the 
hadron's transverse momentum broadening 
in DIS on a nuclear target compared to
a proton or deuteron target \cite{Kopeliovich}.
Indeed, when a hadron is observed in the final state, neither the
quark nor the prehadron could have had inelastic scatterings. The
prehadron-nucleon elastic 
cross section is very small compared to the quark cross section, 
so that the hadron's $p_T$-broadening
originates dominantly during parton propagation. As shown in
\cite{ptbroad,ptbroad2}, the quark's momentum
broadening $\Delta p_T^2$ is proportional to the quark path-length
in the nucleus. If the prehadron formation time 
has the form \eqref{eq:scalingvar} as argued in the last section, we obtain: 
\begin{align*}
  \Delta p_T^2 \propto \vev{t_*} \propto z_h^\lambda (1-z_h) \nu  \ ,
\end{align*}
unless the distance of the quark production point from the surface of
the nucleus is smaller than $\vev{t_*}$. 
Then we should expect a decrease of $\Delta p_T^2$ with increasing
$z_h$. This  would be a clear and model-independent 
signal of in-medium prehadron
formation: indeed, if the quark were traveling
through the whole nucleus before prehadron formation 
$\Delta p_T^2$ would only depend on the nucleus size and not on
$z_h$. 
A related observable is the $z_h$-dependence of the Cronin effect, which is
likewise expected to decrease with increasing $z_h$
\cite{Kopeliovich}.

%Its magnitude can then be related to the overall scale of
%$l_*$ using theoretical models for quark-nucleon interactions. 
%The $\nu$- and $Q^2$-dependence of $\Delta p_T^2$ may then be used to
%pin down the exact functional form of $l_*$.

The CLAS collaboration can perform multi-differential $p_T$-broadening
measurements in all kinematic variables
thanks to a very high beam luminosity. A few preliminary data from
CLAS are already available \cite{Brooks-JLAB}. The HERMES
collaboration is also studying the $p_T$-broadening at a larger beam
energy but with a lower statistics. The scaling analysis proposed in
the previous section will be useful to cross-check the results of these
measurements.

\section{Conclusions and perspectives}
\label{sec:conclusions}

Use of hard processes to probe medium processes in A+A collisions
requires a detailed understanding of the hadronization process, which
can be studied in lepton-nucleus scatterings.
The scaling analysis \cite{Accardi06}
of pion attenuation at HERMES demonstrates for the first time a 
scaling of the hadron attenuation ratio $R_M$ which is  
compatible with a short prehadron formation time of the order or
smaller than the nuclear radius. Thus, it favors prehadron absorption as
dominant mechanism for hadron suppression instead of gluon radiation
off a struck quark. This conclusion will be soon checked by hadron
$p_T$-broadening data. 

Much more can be studied in lepton-nucleus scatterings. (i) In the meson
sector, the suppression of $\eta$ mesons at RHIC, which is of similar
magnitude than for $\pi$, seems to favor long-lived partons
\cite{pieta}. Measuring $\eta$ and heavier meson attenuation at HERMES
and CLAS will check the correctness of such interpretation. (ii)
Understanding the proton anomaly in nDIS will shed light on baryon
transport in nuclear matter and on the baryon anomaly observed in p+A
and A+A collisions. Measurements of $\Lambda$ and other
baryons at HERMES and CLAS will be needed in this respect.
(iii) Neither HERMES nor CLAS are able to study the hadronization of
heavy quarks, because of limited luminosity and limited Bjorken's $x$
coverage, respectively. The proposed Electron-Ion Collider
\cite{Deshpande:2005wd} 
is well suited for such studies, thanks to its low-$x$ coverage and
high luminosity. Study of D and B meson suppression will settle the
single electron puzzle at RHIC and will put interpretation of
LHC data on a firmer ground. Study of ``normal'' $J/\psi$ suppression
will help in distinguishing competing mechanism and in building a
precise baseline for measurements of the ``anomalous'' suppression in
A+A collisions.

\begin{acknowledgement}

I am very grateful to the organizers of this workshop for the
financial support they offered me. I would like to thank 
V.~Muccifora, D.~Gr\"unewald, H.~J.~Pirner for their
collaboration on many results discussed in this work, and  
J.~W.~Qiu, P.~Di Nezza, F.~Arleo, C.~Salgado, T.~Falter, K.~Gallmeister
for valuable discussions. Figure 1 has been realized using JaxoDraw by
D.~Binosi and L.~Theussl. 
This work is partially funded by the US Department of Energy grant
DE-FG02-87ER40371. 

\end{acknowledgement}

% Non-BibTeX users please use

\end{document}